\documentstyle[multicol,aps,epsf,pra,psfig]{revtex}

\begin{document}
\draft

\newcommand{\ket}[1]{\left | #1 \right \rangle}
\newcommand{\bra}[1]{\left \langle #1 \right |}

\title{ Entanglement quantification and purification in continuous variable systems}
\author{S. Parker, S. Bose, M.B. Plenio}
\address{Optics Section, The Blackett Laboratory, Imperial College, London SW7 2BZ, England}
\date{\today}

\maketitle
\begin{abstract}

We develop theoretical and numerical tools for the quantification of entanglement in systems with continuous degrees of freedom. Continuous variable entanglement swapping is introduced and based on this idea we develop methods of entanglement purification for continuous variable systems. The success of these entanglement purification methods is then assessed using these tools.

\end{abstract}


\begin{multicols}{2}

\section*{introduction}

Recently the theoretical idea of teleportation in systems with 
continuous variables has been developed\cite{telecont,telesque,enk,ralph} and shortly
afterwards demonstrated by Furusawa et.\@ al.\@\cite{teleexp}. The efficiency
of entanglement manipulation protocols critically depends on the
quality of the entanglement that one can generate. It is therefore
essential firstly, to be able to quantify the amount of entanglement in 
systems with continuous variables and secondly, to develop methods 
of entanglement purification that allow the distillation of entanglement 
by local means in such systems. The present paper provides answers 
to both of these essentials.

Most analysis of entanglement in continuous variable systems relies on expressing the state of the system in terms of some discrete but infinite basis, often with mathematical techniques from quantum optics\cite{enk,leonhardt}. In this way the above procedures can be demonstrated and the entanglement quantified. Such analysis is convenient but not necessary for the theoretical description of these processes. However, for quantifying entanglement it is essential.

In pure finite bipartite systems the Schmidt basis of a particular system is useful, particularly because its coefficients allow us to calculate the {\em entropy of entanglement}\cite{benn}. In this paper we will first describe how such a basis occurs in continuous systems from the mathematical area of integral eigenvalue equations and how the entanglement can be calculated from it. We then present two classes of continuous entangled states and calculate their entanglement using a numerical solution from statistical mechanics\cite{sm}.

Purification is the process by which the entanglement in a bipartite state shared between two spatially separated parties (Alice and Bob) can be increased using only local operations and classical communication\cite{new}. There are many methods of achieving this for discrete pure systems such as the Procrustean method\cite{benn} (for one copy of the bipartite state) and using entanglement swapping\cite{sugent} (using two copies of the bipartite state one of which is not shared). So far it has not been shown that any of these procedures have a continuous analogue which is able to purify an entangled state.

Here we will show that a continuous generalization of an entanglement swapping procedure using the continuous controlled-NOT and Hadamard gates introduced by Braunstein\cite{braun} is able to produce purification in one of our two classes of continuous entangled states.

Section \ref{mathss} first presents necessary mathematics from integral eigenvalue equations in the context of continuous quantum mechanical systems. Section \ref{gands} gives two quantum gates and presents the two classes of entangled states. The calculation of the entanglement is attempted analytically as for as possible in section \ref{calcent} and the numerical procedure is presented and used where necessary. In section \ref{purific} entanglement purification is attempted and validated using the techniques developed in the previous section, and finally conclusions are given in section \ref{conclu}.

\section{Preliminary mathematics}
\label{mathss}

In this section we present some of the mathematical tools that are used in this paper for the description of continuous variable systems. For continuous states we can express a pure bipartite system as
\begin{equation}
\label{genstate}
\ket{\psi}_{12} = \int \psi(x,y) \ket{x}_1 \ket{y}_2 \, dx dy,
\end{equation}
where $\ket{x}$ are the eigenstates of the continuous system (position, say). We wish to find the Schmidt basis for the bipartite system for which either partial density operator of the system, e.g.
\begin{eqnarray}
\label{genden}
\rho_1 &=& \int {}_2 \langle x \ket{\psi} {}_{12} \, \, {}_{12} \langle \psi \ket{x}_2 \, dx \nonumber \\
&=& \int \rho_1(x,y) \ket{x} {}_1 \, {}_1 \bra{y} \, dx dy
\end{eqnarray}
is diagonal. All the necessary mathematics is covered in the area of integral eigenvalue equations\cite{tricomi}: so suppose that we wish to find the eigenvalues of the {\em kernel} $\rho_1$, that is, we wish to find $\phi_i(x)$'s such that
\begin{equation}
\label{cee}
\int \rho_1(x,y) \phi_i(y) \, dy = \lambda_i \phi_i(x)_, 
\end{equation}
where $\lambda_i$ is the eigenvalue corresponding to $\phi_i(x)$. For Hermitian kernels (for which $\rho_1^*(y,x) = \rho_1(x,y)$) the eigenvalues are real and the set of eigenfunctions will be linearly independent, complete and also orthogonal
\begin{equation}
\label{orthog}
\int \phi_i(x) \phi_j^*(x) \,dx \ = 0, \qquad i\ne j.
\end{equation}

A special class of kernels are those which are {\em quadratically integrable}:
\begin{equation}
\label{l2def}
\int\!\!\!\int K(x,y) dx\, dy\ \qquad\mbox{converges,} 
\end{equation}
where the ranges may be finite or infinite. These are called {\em $L_2$ kernels} as they are integrable over the $L_2$ space. Some basic properties of Hermitian $L_2$ kernels are:
\begin{itemize}
\item they have infinitely many non-zero eigenvalues or are PG (Pincherle-Goursat) kernels, ones which can be decomposed into the form
\begin{equation}
\label{PGkern}
\rho(x,y) = \sum^n_{i=1} X_i(x) Y_i(y), 
\end{equation}
where $\{X_i(x)\}$ and $\{Y_i(y)\}$ are two linearly independent sets of functions.
\item Their eigenvalues have no accumulation points, i.e.\@ the eigenvalues do not form a continuous set, except at zero.
\item The set of eigenvalues converges in the following ways
\begin{eqnarray}
\label{eigconv}
\sum_{i=1}^{\infty} \lambda_i^n = A_n = Tr(\rho_1^n) \equiv \int \rho_1^n(x,x) \,dx \nonumber \\
 \forall \mbox{ finite n,}
\end{eqnarray}
where 
\begin{equation}
\label{Kton}
\rho_1^2(x,y) = \int \rho_1(x,z) \rho_1(z,y) \, dz\
\end{equation}
and further powers of $\rho_1$ can be obtained by induction.
\item A Hermitian $L_2$ kernel can be written as the expansion
\begin{equation}
\label{expkern}
\rho_1(x,y) = \sum_{i=1}^{\infty} \lambda_i \phi_i(x) \phi_i^*(y)
\end{equation}
(where the $\phi_i(x)$ are normalised to the square modulus) or the kernel is PG when there are only $n$ terms, with $n$ as in equation (\ref{PGkern}). 
\end{itemize}

We see from the above that for Hermitian $L_2$ kernels we can {\em always} find a diagonal Schmidt decomposition into an orthogonal basis given by the kernel's eigenfunctions. The dimension of this basis therefore depends on the number of linearly independent eigenfunctions the kernel possesses. 

The measure of entanglement for a pure bipartite system is now easily generalized to continuous variables and is just the Von Neumann entropy of either partial density operator of the system\cite{benn}
\begin{equation}
\label{defnE}
E(\rho_{12}) \equiv S(\rho_1) = S(\rho_2) = -\sum_i \lambda_i log_2 \lambda_i,
\end{equation}
the number of terms in the sum depending on the form of $\rho_1$ or $\rho_2$. We will call this the {\em entropy of entanglement} or just the {\em entanglement}.

Properties such as concavity, subadditivity, strong subadditivity and the triangle inequality also follow for this measure of entanglement as they do its finite partner\cite{wehrl} provided the relevant quantities converge when we deal with infinite systems. The invariance under unitary transform of the subsystems can also be proved, where the transform is of the form
\begin{equation}
\label{formofU}
U\ket{x} = \int U(y,x) \ket{y} \,dy\
\end{equation}
with
\begin{equation}
\label{propofU}
U^{\dagger}U\ket{x} = \ket{x}  \Rightarrow \int U^*(y,z)U(y,x) \,dy \ = \delta(x-z) .
\end{equation}
Transforming the subsystems independently 
\begin{equation}
\label{transindep}
\rho_{12}' = U_1U_2\rho_{12}U^{\dagger}_1U^{\dagger}_2
\end{equation}
and tracing out system 2 gives
\begin{equation}
\label{tracout}
\rho_{1}'  =  U_1 \rho_1 U^{\dagger}_1,
\end{equation}
as the trace is invariant under unitary transform. It can then be shown that $\rho'_1$ has the same eigenvalues as $\rho_1$
\begin{eqnarray}
\label{ee}
\rho_1 \ket{\phi_i}_1 &=& \lambda_i \ket{\phi_i}_1 \qquad \mbox{or} \nonumber \\
\int \rho_1(x,y) \phi_i(y) \, dy \ &=& \lambda_i \phi_i(x) 
\end{eqnarray}
but its eigenfunctions are related to those of $\rho_1$ by
\begin{equation}
\label{eigrel}
\phi_i(x) = \int U^*_1(y,x)\phi'_i(y) \, dy.
\end{equation}

These mathematical techniques, particularly the latter method of showing the equivalence of eigenvalue equations by transforming the eigenfunctions, will be used in the next sections to determine the amount entanglement in continuous variable systems.

\section{Quantum gates and classes of entangled states}
\label{gands}

We now turn to some classes of entangled states in continuous systems, which will be a convenient point to introduce two quantum gates generalized from discrete gates\cite{braun}. The first is the continuous Hadamard transform which is in fact just the Fourier transform, and in which we will include the scale length, $\sigma$, explicitly:
\begin{equation}
\label{four}
 {\cal F}  \ket{x} = \frac{1}{\sqrt{\pi}\sigma} \int e^{2ixy / \sigma^2} \ket{y} \,dy \ .
\end{equation}
This is also, of course, the transform used to go from the position to the momentum basis if we set $\sigma = 1$ and work in units $\hbar = 1/2$. The inverse, ${\cal F}^{\dagger}$ is obtained by replacing $i$ by $-i$ giving the result that ${\cal F F}^{\dagger} \ket{x} = {\cal F}^{\dagger} {\cal F} \ket{x} = \ket{x}$. Note that the scale length, $\sigma$, is normally inserted to make the expression in the exponential dimensionless but we will include it as a convenient scale with which to compare various lengths in the states we are about to form.

The second important gate is the two particle controlled-NOT gate:
\begin{equation}
\label{cnot}
 {\cal C}_{12} \ket{x}_1 \ket{y}_2 = \ket{x}_1 \ket{y+x}_2 ,
\end{equation}
which is not its own inverse. This is obtained by replacing the $+$ with a $-$ sign on the right hand side.

Another version of the controlled-NOT, which is its own inverse, is 
\begin{equation}
 {\cal C}' \ket{x,y} = \ket{\frac{x+y}{\sqrt{2}},\frac{x-y}{\sqrt{2}}} .
\end{equation}
This is a more symmetric gate and is in fact the transform produced by a 50:50 beam splitter on the quadrature wavefunctions in quantum optics\cite{leonhardt}. However, we will proceed with the original definition of the CNOT for simplicity (primarily of mathematics).

We can now define the `entangling' operation and its inverse:
\begin{equation}
\label{entgate}
{\cal E}_{12} = {\cal C}_{12} {\cal F}_1 \qquad {\cal E}^{\dagger}_{12} = {\cal F}^{\dagger}_1 {\cal C}^{\dagger}_{12},
\end{equation}
and form our first class of entangled states by applying ${\cal E}$ to two Gaussian wavepackets (aside from normalisation): 
\begin{equation}
\label{gauss1}
\ket{G_{\alpha}(x_1)}_1 = \int exp \left[ -\frac{(x-x_1)^2}{\alpha^2\sigma^2} \right]\ket{x}_1 \, dx
\end{equation}
and $\ket{G_{\beta}(x_2)}_2$, resulting in the state
\begin{eqnarray}
\label{gausent}
  \begin{cal} C \end{cal}_{12} \begin{cal} F \end{cal} _1 && \ket{G_{\alpha}(x_1)}_1 \ket{G_{\beta}(x_2)}_2 \nonumber \\ 
 &&=  \int  exp\left[\frac{1}{\sigma^2}\left(-x^2\alpha^2-\frac{y^2}{\beta^2} +2ix_1x\right)\right] \nonumber \\
&&\qquad \quad \ket{x}_1 \ket{x+y+x_2} \, dxdy \ \nonumber \\ 
&& \equiv  \ket{B_{\alpha\beta}(x_1,x_2)}_{12} .
\end{eqnarray}

Such states can be used to demonstrate teleportation of an unknown state, the fidelity of the teleportation increasing as $\alpha$ and $\beta \rightarrow 0$, where the state becomes like an infinitely squeezed two mode squeezed state or an EPR state\cite{vaidman}. We will call the states $\ket{B_{\alpha\beta}(x_1,x_2)}_{12}$ {\em partially correlated entangled states}. 

The second class of states we will call {\em two-mode cat states}\cite{sanders,gerry}, as they are states which are like two Schr\"odinger cat states whose locations are correlated with each other quantum mechanically:
\begin{eqnarray}
\label{cat}
\ket{C(d)}_{12}& = &\int \left[A_0 e^{-\left(x-d\right)^2 - \left(y+d\right)^2}  + A_1 e^{-\left(x+d\right)^2 - \left(y-d\right)^2} \right] \nonumber \\
&& \qquad \ket{x}_1 \ket{y}_2\, dx dy \nonumber \\
& = & \int \sum_{j=0}^1 \left[A_j e^{-\left(x-(-1)^j d\right)^2 - \left(y+(-1)^jd\right)^2} \right] \nonumber \\
&& \qquad \ket{x}_1 \ket{y}_2 \, dxdy. 
\end{eqnarray}
The complex coefficients, $A_j$, are such that $|A_0|^2 + |A_1|^2 = 1$ (so the state is not normalised correctly). This state is a superposition of the first particle being located around $d$ and the second around $-d$ and {\em vice versa} and so is not of great use in teleportation. The scale length does not appear here (it is set to unity) as an increase in scale length is equivalent to a decrease in the value of $d$.

It is now natural to ask what the amount of entanglement in the states $\ket{B_{\alpha\beta}(x_1,x_2)}_{12}$ and $\ket{C(d)}_{12}$ is, which is the aim of the next section.

\section{Quantification of the entanglement}
\label{calcent}
\subsection{Mathematics}

We will now proceed in calculating the entanglement of the first class of states, $\ket{B_{\alpha\beta}(x_1,x_2)}_{12}$. We have already shown that the entanglement should be equal when calculated using either partial density matrix (from the fact that a Schmidt basis exists) but we will show this explicitly by showing that the eigenvalue equations they give can both be transformed into the {\em same} eigenvalue equation. The two eigenvalue equations are
\begin{eqnarray}
\label{rho1again}
&&\int exp\biggl[\frac{1}{\sigma^2}\biggl(-\left(\alpha^2+ \frac{1}{2\beta^2}\right)(x^2+x'^2) \nonumber \\
&&+2ix_1(x-x')+\frac{xx'}{\beta^2} \biggl) \biggl]\phi_i^{(1)}(x')  \,dx' \ = \lambda_i^{(1)} \phi_i^{(1)}(x)
\end{eqnarray}
\begin{eqnarray}
\label{rho2again}
&&\int exp\biggl[\frac{1}{\sigma^2} \biggl(\left(\frac{1}{2(\alpha^2\beta^2+1)}-1\right)  \frac{y^2+y'^2}{\beta^2} \nonumber \\
&&+\frac{yy'}{\beta^2(\beta^2\alpha^2+1)} \biggl) \biggl]\phi_i^{(2)}(y')  \,dy' \ = \lambda_i^{(2)} \phi_i^{(2)}(x).
\end{eqnarray}
We can now recast the eigenfunctions of equation (\ref{rho1again}):
\begin{equation}
\label{recast}
\phi_i^{(1)}(x') \rightarrow \phi_i^{(1)}(x') exp(-2ix_1x')
\end{equation}
thereby absorbing the complex exponential into the eigenfunctions. Next we perform the following change of variables in equation (\ref{rho1again})
\begin{equation}
\label{chgvar1}
x \rightarrow \sqrt{2}x\beta\sigma \qquad x' \rightarrow\sqrt{2}x'\beta\sigma
\end{equation}
and in equation (\ref{rho2again})
\begin{equation}
\label{chgvar2}
x \rightarrow x\sqrt{2(\alpha^2\beta^2+1)} \beta\sigma \qquad x' \rightarrow \sqrt{2(\alpha^2\beta^2+1)}x'\beta\sigma .
\end{equation}
These changes of variables are the same for both dashed and undashed variables so this change can be absorbed into the eigenfunctions. All these changes leave the set of eigenvalues unchanged and give the same eigenvalue equation:
\begin{eqnarray}
\label{mainee}
\int^{\infty}_{-\infty} \underbrace{exp\left[-\left(1+2\alpha^2\beta^2\right) (x^2+x'^2) + 2xx' \right]}_{K(x,x')} && \phi(x')  \,dx' \nonumber \\
 =&& \lambda \phi(x)
\end{eqnarray}
where $K(x,x')$ is the Kernel of our integral eigenvalue equation. Its trace is
\begin{equation}
\label{tracemain}
\int K(x,x) \, dx = \frac{\sqrt{\pi}}{2\alpha\beta} . 
\end{equation}
The set of eigenfunctions are of course independent of any normalisation constant but we will find that this constant is our primary check for the convergence of the numerical solution that follows in that the eigenvalues should sum to this constant as in equation (\ref{eigconv}). 

More importantly we notice that the eigenvalues are independent of the scale length, $\sigma$, and the value of $x_1$ or $x_2$. It is only dependent on the product of widths of the original Gaussian distributions with respect to $\sigma$.  

The two particle cat state has partial density operator (aside from normalisation)
\begin{eqnarray}
\label{catden1}
\rho_1 = \int && \sum_{j=0}^1 \sum_{k=0}^1 \left[A_j A_k^* e^{-\left(x-(-1)^j d\right)^2 - \left(x'-(-1)^k d\right)^2 + 2d^2 \delta_{jk}} \right] \nonumber \\
 && \ket{x}_1\bra{x'} \, dxdx' ,
\end{eqnarray}
which is not easily transformed into a simpler form. Its trace is
\begin{equation}
\label{cattr1}
Tr(\rho_1) = \sqrt{\frac{\pi}{2}}\left( e^{2d^2} + (A_0 A_1^* + A_0^* A_1)e^{-2d^2} \right).
\end{equation}
This, again, will be our primary check for convergence of the numerical model that follows.

\subsection{Numerical procedure for partially correlated states}

We cannot solve many of the integral eigenvalue equations we encounter so we must use some numerical approximation\cite{sm}. The most direct approach is to solve a discrete eigenvalue equation by approximating the integral by the rectangle rule. Our eigenvalue equation has infinite limits so there must also be a  cut-off point in the limits beyond which we do not approximate the integral. 

First, therefore, we discretize the eigenvalue equation (\ref{mainee}) into $2n+1$ parts ($i=-n \ldots, 0, \ldots n$) each of width $\delta$ covering the range $-w \le x \le w$ where $w = n\delta$. Our eigenvalue equation then becomes
\begin{equation}
\label{dismainee}
\delta \sum_{p=-n}^n \rho_{pq} \phi_p = \lambda \phi_q,
\end{equation}
where the indices now denote particular matrix and vector elements (rather than particular eigenvectors) and for the Bell states
\begin{equation}
\label{disrho}
\rho_{pq} = exp\left[\left(-(1+2\alpha^2\beta^2)(p^2+q^2)+2pq\right)\delta^2 \right] .
\end{equation}
Our entanglement measure is then
\begin{equation}
\label{disE}
E(\alpha,\beta) = \sum_r \left(\frac{\lambda_r}{\sum_s \lambda_s} log_2 \left(\frac{\lambda_r}{\sum_t \lambda_t}\right)\right),
\end{equation}
where the outer sum is over the set of eigenvalues and the sums over $s$ and $t$ are to normalise the set of eigenvalues.

As mentioned above the primary check for the convergence of this solution is to compare $\sum \lambda$ with equation (\ref{tracemain}). We have two independent parameters that we may vary (at each value of $\alpha$ and $\beta$) out of the three related parameters $n$, $\delta$ and $w$. In practice for most values of $\alpha$ and $\beta$, $2n+1 = 201$ and $w$ around 10 standard deviations from the mean were sufficient for this sum to be equal to the trace accurate to 5 significant figures, this accuracy becoming greater with increased $n$ and decreased $\delta$.

We can see that what we are effectively doing in this procedure is sampling the spectrum of eigenvalues of the density operator over discrete ranges of width $\delta$ by modelling the continuous system as a discrete $2n+1$ level system and taking the limit $n \rightarrow \infty$. With the above convergence of the model we can be confident that both an adequate range of the integral has been sampled and that it has been sampled to an adequate precision.

\begin{figure}
\begin{center} 
\leavevmode 
\centerline{\hbox{\psfig{figure=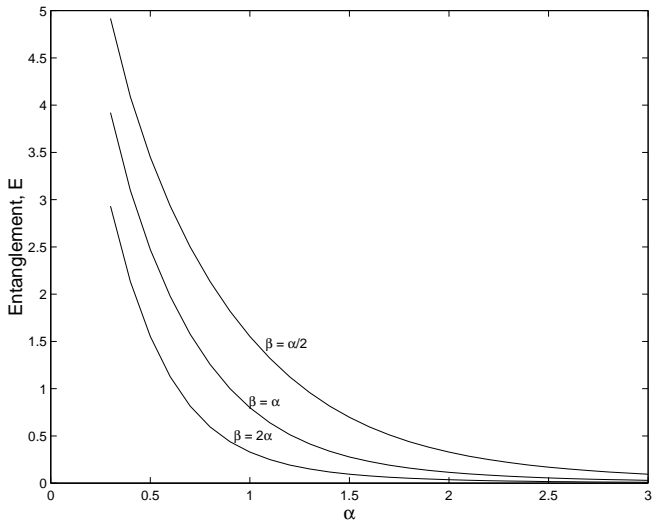,height=7.cm}}}
\end{center}
\caption{\narrowtext Entanglement of a partially correlated state in terms of $\alpha$ and $\beta$, the widths of the Gaussians from which they are formed.}
\label{Bellres}
\end{figure}

We used numerical procedures for eigenvalue problems from the NAG library to solve this problem for varying values of $\alpha$ and $\beta$. The results are shown in figure \ref{Bellres}.

As we expect the entanglement is increased when the parameters $\alpha$ and $\beta$ are reduced, becoming infinite as these parameters approach zero, as will be shown in the next section. At this point it becomes harder to see convergence in the numerical procedure.

\subsection{Analytical results for partially correlated states}

We can now be even more general than this:\@ we see that the form of the eigenvalue equation (\ref{mainee}) is
\begin{eqnarray}
\label{maineeform}
\int^{\infty}_{-\infty} exp\left[-(1+P) (x^2+x'^2) + 2xx' \right]&&\phi(x')  \,dx' \ \nonumber \\ 
 = &&\lambda \phi(x)
\end{eqnarray}
with the entanglement being decreasing with increase in the parameter $P$. In fact, a state of the form
\begin{equation}
\label{pargen}
\int exp \left( -a x^2 -b y^2 + 2c x y + d x + e y \right) \ket{x} \ket{y} \, dx dy
\end{equation}
has a parameter 
\begin{equation}
\label{parrr}
P = 2\left(\frac{ab}{c^2} - 1\right)
\end{equation}
independent of $d$ and $e$, which can be shown by making appropriate linear changes of variables  $ x \rightarrow Ax + B $ and $ x' \rightarrow Ax' + B$ in the eigenvalue equation formed by the density operators of the state (\ref{pargen}).

For a two mode squeezed state with squeezing parameter $r$
\begin{equation}
\label{squeeze}
\psi(x,y) = exp \left( -\frac{1}{4} \left(e^{2r} (x+y)^2 + e^{-2r} (x-y)^2 \right) \right)
\end{equation} 
this parameter is 
\begin{equation}
\label{squP}
P = 2cosech^2(2r).
\end{equation}
Such a state can, of course, be written analytically in the number basis\cite{enk} 
\begin{equation}
\label{squenk}
\ket{\psi}_{12} = \frac{1}{cosh(r)} \sum_{n=0}^{\infty}(tanh(r))^n \ket{n}_1 \ket{n}_2
\end{equation}
which {\em is} the Schmidt basis for this state and from this the entanglement can be calculated as
\begin{equation}
\label{Eenk}
E = cosh^2(r) \, log_2 (cosh^2(r)) -  sinh^2(r) \, log_2 (sinh^2(r)).
\end{equation}
It is reassuring to note that the above simulation can reproduce this analytical result accurate to 6 significant figures. More importantly we now have an analytical result for the entanglement of the partially correlated states which is equation (\ref{Eenk}) with the substitution $2r = arcsinh(1/\alpha\beta)$, which follows directly from equation (\ref{squP}).

\subsection{Numerical procedure for two-mode cat state}

\begin{figure}
\begin{center} 
\leavevmode 
\centerline{\hbox{\psfig{figure=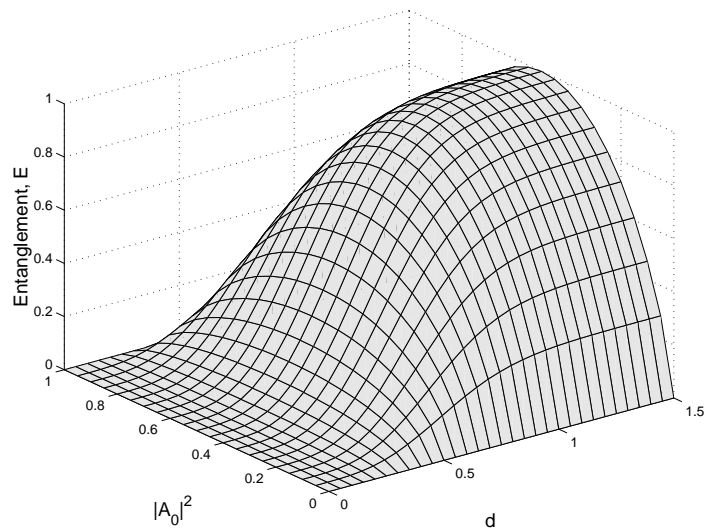,height=7.cm}}}
\end{center}
\caption{\narrowtext Entanglement of the cat states against (half) the distance between Gaussians, $d$, and the coefficient $A_0$. The entanglement is greatest for high values of $d$ where the Gaussians become orthogonal, and for $|A_0|^2 = 0.5$ as with discrete entanglement.}
\label{catres}
\end{figure}

Now we move on to the entanglement of the cat states. The eigenvalue equation for these cannot be rewritten in terms of the parameter, $P$, so we proceed directly by discretizing the density operator of equation (\ref{catden1}) and making it the kernel of equation (\ref{dismainee}). Results for the entanglement of the cat states are shown in figure \ref{catres} against the parameters $d$ and $|A_0|^2$.

Note that the entanglement for any given value of $d$ is maximum when $|A_0|^2 = 1/2$ and that the entanglement increases with $d$ for given values of $A_0$, approaching the limit $E = -|A_0|^2 log_2 |A_0|^2 - |A_1|^2 log_2 |A_1|^2$ as $d \rightarrow \infty$, where the separated Gaussians become orthogonal. Note also that only two eigenvalues dominated the contribution to the entropy, although we do not expect these states to be written in some basis as a two level system. These states, therefore, behave very much like discrete 2-level entangled systems.
 
\section{Purification}
\label{purific}

Now that we have a reliable method of calculating the entanglement we will move on to attempt entanglement purification in the two classes of states.  The first point to note is that purification is always possible for such states as we can just make a measurement, one of the results of which is a projection onto the two levels of the Schmidt basis with the largest Schmidt coefficients. After this we can then perform discrete purification to obtain highly entangled two level states which may have higher entanglement that our original continuous state. However we wish to work more in the spirit of continuous systems using continuous operations and producing continuous entangled states.

\begin{figure}
\begin{center} 
\leavevmode 
\centerline{\hbox{\psfig{figure=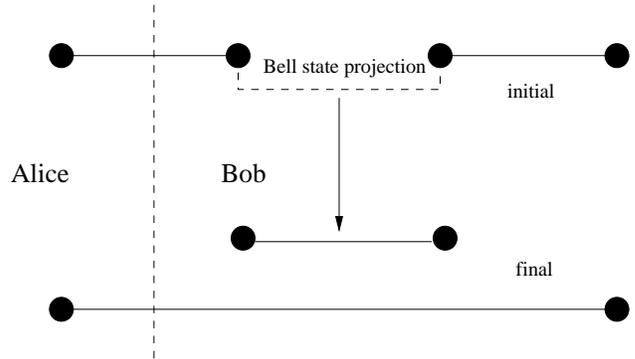,height=4.8cm}}}
\end{center}
\caption{\narrowtext The entanglement swapping procedure. Bob, holding a copy of the entangled state shared by himself and Alice, performs a Bell state measurement on a particle from each pair and, for certain measurement outcomes, the entanglement of the final shared pair is higher than that of the initial shared pair.}
\label{swapsies}
\end{figure}

Again we generalize a purification procedure from discrete systems. The one we have chosen is purification by entanglement swapping\cite{sugent}. In the discrete case Alice and Bob share an entangled state and Bob holds another copy of the same entangled state as in the upper part of figure \ref{swapsies}. Bob performs entanglement swapping by making a Bell state projection on one particle from each pair of entangled particles. This leaves the remaining two particles (one on Bob's side and one on Alice's) in an entangled state. If the original entangled state were maximally entangled then the final pair will also be maximally entangled. This is the standard entanglement swapping procedure\cite{sugpur}. If the original state was less than maximally entangled then the entanglement of the final pair will, for certain measurement outcomes, be less entangled than the original states, but for the remaining outcomes, will be maximally entangled. This procedure is, therefore, probabilistic as there is dependence on the outcome of these measurements. Entanglement swapping has also been recently achieved independently in continuous variable systems\cite{swappaper}. 

\subsection{Imperfectly correlated states}

We will attempt the analogous procedure here except that the Bell state measurement will be replaced by a reverse entangling operation ${\cal E}^{\dagger}$ and projective infinite resolution measurements on the two particles.

The procedure for the partially correlated Bell state is
\begin{eqnarray}
\label{Bellswap}
& & {}_2\bra{a} {}_4 \bra{b} \left( {\cal E}^{\dagger}_{24}  \left(\ket{B_{\alpha\alpha}(0,c)}_{12} \ket{B_{\beta\beta}(0,c)}_{34} \right) \right) \nonumber \\
& = & \int exp\biggl[\frac{1}{\sigma^2(\alpha^2+\beta^2)}\Bigl(-x^2 g(\alpha,\beta) - y^2 g(\beta,\alpha)+ 2xy \nonumber \\
&& \quad \qquad+ 2b(y-x) - 2ia(y\alpha^2+x\beta^2)\Bigl)\biggl] \ket{x}_1\ket{y}_3
\end{eqnarray}
where
\begin{equation}
\label{defineg}
g(\alpha,\beta) = \alpha^4 +\alpha^2\beta^2 +1 .
\end{equation}
Has the entanglement increased? First notice that from equation (\ref{parrr}) the entanglement does not depend on $a$ or $b$ and so is not probabilistic. This indicates that the entanglement cannot have increased otherwise we would have deterministic purification. Indeed this is true as the parameter for this state is $P_{swap} = 2[ (\alpha^4 +\alpha^2\beta^2+1) (\beta^4 + \beta^2\alpha^2 +1)  -1]$ whereas before the swapping process the parameter was $P_0 = 2\alpha^4 $ or $ 2\beta^4$. But $P_{swap} \ge P_0$ in either case and the entanglement is strictly decreasing with increase in $P$ so the swapped pair has less entanglement.

Of course, our final projections in this method were onto the unphysical states $\ket{a}$ and $\ket{b}$ but further calculations indicate that with finite width projections the parameter $P$ still increases. Such calculations involve $6^{th}$ degree polynomials in the width parameters ($\alpha$ and $\beta$ etc.\@) so proving that $P$ increases for all values of these parameters is difficult and we have not been able to do so analytically. However, graphical results indicate that this is true.

\subsection{Two-mode cat states}

We now attempt a similar method with the two-mode cat states, but setting the scale length $\sigma = 1$ and making {\em finite} resolution measurements:
\begin{eqnarray}
\label{catswappp}
 \ket{\psi}_{14} &=& {}_2\bra{G_{\mu}(a)} {}_3 \bra{G_{\mu}(b)} \left( {\cal E}^{\dagger}_{23} \left( \ket{C(d)}_{12} \ket{C(d)}_{34} \right) \right) \nonumber \\
&=& \int  \sum_{j,k = 0}^1 A_j A_k e^{-(x-(-1)^j d)^2 -(y+(-1)^k d)^2} \nonumber \\
&\times&e ^{\left(  dbh(\mu)   
\left((-1)^j+(1+\mu^2)(-1)^k\right) + 2d^2\delta_{jk} \right)} \nonumber \\
&\times&e ^{\left(  iadh(\mu)  
\left((1+\mu^2)(-1)^j -(-1)^k\right)  \right)} \ket{x}_1 \ket{y}_4 \, dxdy
\end{eqnarray}
where
\begin{equation}
\label{h}
h(\mu) = \frac{2}{2+2\mu^2+\mu^4}.
\end{equation}

Writing this state out in full in the high precision measurement limit, $\mu = 0$
\begin{eqnarray}
\label{blur}
\ket{\psi}_{14} = \int  \Bigl(
 A_0 A_0 \,\, &e^{-2d^2 +2db}& \,\, e^{-(x-d)^2 - (y+d)^2}  \nonumber \\
+A_0 A_1 \,\, &e^{2iad}      & \,\,e^{-(x-d)^2 - (y-d)^2}  \nonumber \\
+A_1 A_0 \,\, &e^{-2iad}     & \,\, e^{-(x+d)^2 - (y+d)^2}  \nonumber \\
+A_1 A_1 \,\, &e^{-2d^2 -2db}& \,\, e^{-(x+d)^2 - (y-d)^2} \Bigl) \nonumber \\ 
&& \ket{x}_1 \ket{y}_4 \, dxdy
\end{eqnarray}
and looking at the particular case where the probabilistic measured values are $a=b=0$ we can see purification for high values of $d$ as the middle two terms now dominate and have equal coefficients. As $d \rightarrow \infty$ they become maximally entangled. This is again very much like the action of discrete entanglement under purification procedures: the coefficients of the states have changed not the states themselves.

\begin{figure}
\begin{center} 
\leavevmode 
\centerline{\hbox{\psfig{figure=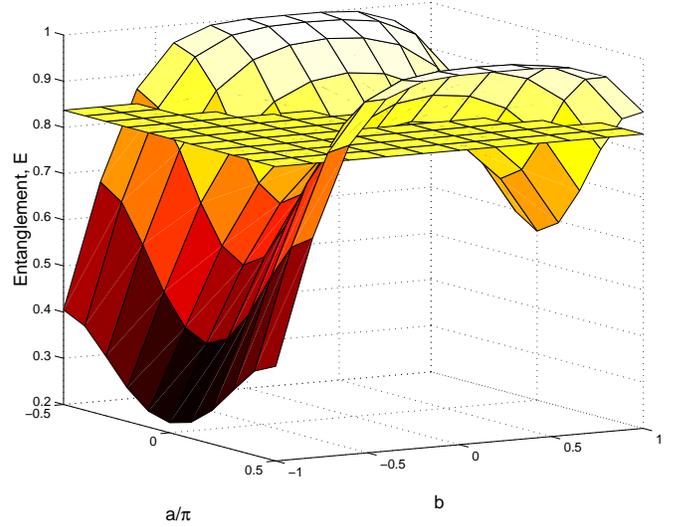,height=7.cm}}}
\end{center}
\caption{\narrowtext Entanglement of swapped cat states with $\mu = 0$. Above the level of the plane purification has been achieved.} \label{catpurr1}
\end{figure}

\begin{figure}
\begin{center} 
\leavevmode 
\centerline{\hbox{\psfig{figure=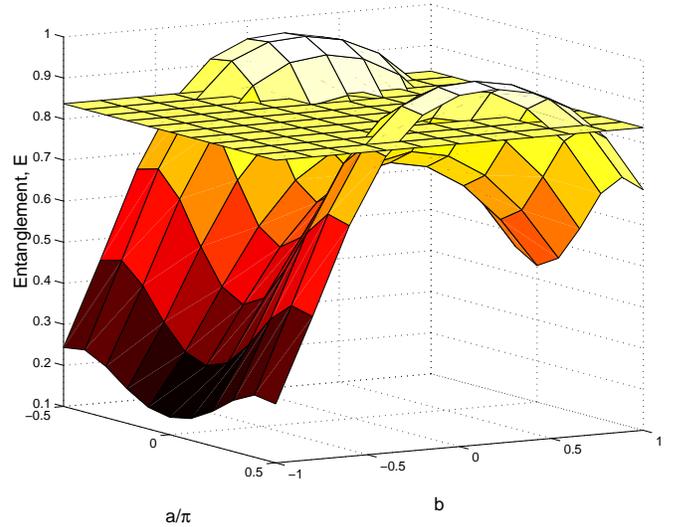,height=7.cm}}}
\end{center}
\caption{\narrowtext Entanglement of swapped cat states with $\mu = 0.5$. Again purification has been achieved above the level of the plane, but with the inaccuracy in the measurement part of the entanglement swapping process the amount of purification is reduced.} \label{catpurr2}
\end{figure}

For the results of figure \ref{catpurr1} and \ref{catpurr2} we have chosen the values $A_0 = \sqrt{0.3}$ and $d = 1.0$. They show the entanglement of the resulting state (\ref{catswappp}) for a range of values of $a$ and $b$ with $\mu = 0$ and $0.5$ respectively.

The horizontal planes are at the level of entanglement of either cat state before the purification procedure is performed, that is, above this plane purification has been achieved. Notice that purification is achieved above the level of entanglement of the initial cat state of equation (\ref{cat}) with parameters 
$\underline{A_0 = \sqrt{0.5}}$ and $d = 1.0$ for which $E = 0.881$.

\section{Conclusions}
\label{conclu}

We have shown that the usual measure of entanglement for pure bipartite discrete systems is easily generalized to continuous systems with knowledge of the mathematical area of integral eigenvalue equations, in particular that a Schmidt basis exists provided the state satisfies certain reasonable conditions of integrability. The calculation of the entanglement, however, is often difficult so a numerical procedure was necessary. The results of the numerical procedure corresponded very well with those of analytical results where these were available and enabled us to test a purification procedure, the entanglement swapping procedure, for the two classes of states we presented. 

Curiously this procedure only succeeded for one of the classes of states, the {\em two-mode cat states} whose characteristics are very much like those of discrete systems, although there does not appear to be a discrete and finite basis in which the states can be written. The purification procedure also acted in a similar manner to the analogous discrete procedure.

For the other class of states, the {\em partially correlated states}, no continuous procedure could be found that increased the entanglement in the state, indeed, no procedure was found where the entanglement of the final state was in any way dependent on the measurement results during the procedure, reassuring us that no purification would be possible. However, we plan to study this problem more carefully.  

What exactly the key difference is between these two types of states which allows purification in one class but not the other is still unclear. There are obvious correspondences between the form of entanglement in the two mode cat states and discrete systems and it would be interesting to find a condition for continuous variable purification, as it has been attempted here, which a state undergoing purification must obey. The fact that purification has been demonstrated, however, in continuous systems is an interesting result.

\section*{Acknowledgements}

This work is supported by the United Kingdom Engineering and Physical Sciences Research Council (EPSRC), the Inlaks Foundation, The Leverhulme Trust, the EU TMR-networks ERB 4061PL95-1412 and ERB FMRXCT96-0066 and the European Science Foundation.

\end{multicols}  

\end{document}